\documentstyle[aps,preprint,tighten]{revtex} 
\begin{document}
\preprint{imsc/98/04/14}
\draft
\title{Gauge field copies} 
\author{Pushan Majumdar\thanks{e-mail:pushan@imsc.ernet.in}\and H.S.Sharatchandra
\thanks{e-mail:sharat@imsc.ernet.in}} 
\address{Institute of Mathematical Sciences,C.I.T campus Taramani.  
Madras 600-113}
\maketitle 
\begin{abstract}
The problem of Wu-Yang ambiguities in 3 dimensions is related to the problem of 
existence of torsion free driebeins for an arbitrary potential.
The ambiguity is only at the level of boundary conditions. We also find that in 
3 dimensions, any smooth Yang-Mills field tensor can be uniquely written as the 
non-Abelian magnetic field of a smooth Yang-Mills potential. 
\end{abstract} 

Wu and Yang \cite{[1]} gave an explicit example of
two (gauge inequivalent) Yang-Mills potentials
$\vec{A_{i}}(x)=\{A^{a}_{i}(x), a=1,2,3\}$ generating the same
non-Abelian magnetic field 
\begin{equation}\label{def2} 
\vec{B}_{i}[A](x)=\epsilon_{ijk} (\partial_{j}\vec{A}_{k} +
\frac{1}{2}\vec{A}_{j}\times\vec{A}_{k}).  
\end{equation} 
Since then
there has been a wide discussion of the phenomenon in the literature
\cite{[2],[3],[4],[5],[6],[8],[7],[9],[10],[11],[12],[13],[14],[18]}. We 
may refer to gauge potentials giving the same
non-Abelian magnetic field, as gauge field copies in contrast to gauge
equivalent potentials which generate magnetic fields related by a
homogeneous gauge transformation. 
If we require all higher covariant
derivatives of $B^{a}_{i}$ also match then there are effectively no gauge
copies \cite{[11]}.

Deser and Wilczek \cite{[4]} first pointed out the consistency condition 
for ${\vec A}_{\mu}$ and ${\vec A}^{\prime}\!_{\mu}={\vec A}_{\mu}+{\vec \Delta}_{\mu}$ 
to generate the same field strength. Using the Bianchi identity, they obtained
that ${\vec \Delta}_{\mu}$ had to satisfy the equation 
\begin{equation}
[\tilde{\bf F}_{\mu\nu}\,,\Delta_{\nu}]=0.
\end{equation}
where in 2 dimensions,
\begin{equation}
\tilde{F}^{\mu\nu\,ab}=\frac{1}{2}\epsilon^{\mu\nu}
\epsilon^{abc}F^{c}\!_{\mu\nu}=M^{a\,b},
\end{equation}
and in 4 dimensions
\begin{equation}
\tilde{F}^{\mu\nu\,ab}=\frac{1}{2}\epsilon^{\mu\nu\rho\sigma}
\epsilon^{abc}F^{c}\!_{\rho\sigma}=M^{a\mu\,,b\nu}.
\end{equation}
Treating this as an eigenvalue equation for $\Delta$, we have the condition 
for existence of  non-trivial solutions of $\Delta$ is that the determinant 
of {\bf M} is zero.
In 2 dimensions the determinant corresponding to ${\bf M}$ vanishes 
identically and there $\Delta$ necessarily has non-trivial solutions.
However in 4 dimensions this determinant in generically non-zero and
there are hardly any gauge copies.
 
This sort of analysis exists only in even dimensions. 
In 3 Euclidean dimensions, we only get the constraint
${\vec B}_i[A] \times {\vec \Delta}_i=0$. This equation has many
solutions, but this is only a consisitency condition. It does not mean that 
any ${\vec \Delta}_i$ satisfying this equation gives a gauge copy.
 Recently Freedman and Khuri
\cite{[18]} have exhibited several examples of continuous families of gauge 
field copies in d=3. Their technique was to use a local map of the gauge 
field system into a spatial geometry with a second rank symmetric tensor 
$G_{ij}=B^{a}_{i}B^{a}_{j}\:detB$ and a connection with torsion 
constructed from it. 

We adopt a different method and directly ask the question as to how many 
different solutions (if any), does the system of equations defined by 
(\ref{def2}) have for any specified ${\vec B}_i(x)$. 
For that we proceed with the analysis 
using the Cauchy - Kowalevsky existence theorems on systems of partial 
differential equations. The equations for the gauge field 
copies are not a priori in the form where this theorem can be applied. 
However by reorganizing the equations a bit they can be brought to 
the form so that these theorems can be applied to that system.

\section{Existence of A for arbitrary B}

Let us first state the Cauchy - Kowalevsky existence theorem which we use
\cite{CK}.

Let a set of partial equations be given in the form
\begin{equation}
\frac{\partial z_i}{\partial x_1} = \sum^{m}_{j=1}\:\sum^{n}_{r=2}
\: G_{ijr}\frac{\partial z_j}{\partial x_r} + G_i
\end{equation}
for values $i=1,...,m $, being $m$ equations in $m$ dependent 
variables ; the coefficients $G_{ijr}$ and the quantities $G_i$ are 
functions of all the variables, dependent and independent. Let $c_1 , ...,
c_m, a_1,...,a_n $ be a set of values of $z_1, ...z_m, x_1,...,x_n$ 
respectively, in the vicinity of which all the functions $G_{ijr}$ and 
$G_i$ are regular ; and let $\phi_1 , ..., \phi_m$ be a be a set of 
functions of $x_2 ,...,x_n$, which acquire values $c_1, ..., c_m $ 
respectively when $x_2 = a_2 , ..., x_n = a_n$ , which are regular in 
the vicinity of these values of $x_2, ..., x_n $, and which are 
otherwise arbitrary. Then {\em a system of integrals of the equations 
can be determined, which are regular functions of } $x_1 , ..., x_n $
{\em in the vicinity of the values } $x_1 = a_1 , x_2 = a_2 , ..., x_n = 
a_n $, {\em and which acquire the values } $\phi_1 ,..., \phi_m $ {\em 
when } $x_1 = a_1$ {\em ; moreover, the system of integrals determined 
in accordance with these conditions, is the only system of integrals 
that can be determined as regular functions. }

Our system of equations is 
\begin{eqnarray}\label{arr1}
{\vec B}_1 &=& \partial_2 {\vec A}_3 - \partial_3 {\vec A}_2 + {\vec A}_2 \times 
{\vec A}_3 \\ 
{\vec B}_2 &=& \partial_3 {\vec A}_1 - \partial_1 {\vec A}_3 + {\vec A}_3 \times 
{\vec A}_1 \label{arr11}\\ 
{\vec B}_3 &=& \partial_1 {\vec A}_2 - \partial_2 {\vec A}_1 + {\vec A}_1 \times 
{\vec A}_2 \label{bia}
\end{eqnarray}
where ${\vec B}_1 , {\vec B}_2$ and ${\vec B}_3$ are treated as given variables 
and we want to solve for ${\vec A}_1, {\vec A}_2$ and ${\vec A}_3$. With this 
definition of the $B's$, the bianchi identity $D_i B_i=0$ follows 
automatically. However the  existence theorem cannot be applied directly to 
this set of equations.  For that we rewrite the equations in a different way. 
Consider
\begin{eqnarray}\label{arr2}
\partial_3 {\vec A}_2 &=& \partial_2 {\vec A}_3 + {\vec A}_2 \times {\vec A}_3 - 
{\vec B}_1 \\ 
\partial_3 {\vec A}_1 &=& \partial_1 {\vec A}_3 - {\vec A}_3 \times {\vec A}_1 + 
{\vec B}_2 . \label{ar0}
\end{eqnarray}
The existence theorem implies that we have solution for ${\vec A}_1$ and ${\vec A}_2$
for any specified ${\vec B}_1, {\vec B}_2 $ and ${\vec A}_3 $. But ${\vec A}_1$ and 
${\vec A}_2$ so obtained have to satisfy equation (\ref{bia}). Is this always
possible with
some choice of ${\vec A}_3 $, and if yes, is the choice of ${\vec A}_3 $ unique?
To address this question, we presume that the initial data on $x_3=0$ satisfies 
equations (\ref{arr1})-(\ref{bia}). This is always possible for any given ${\vec
B}_i(x)$ as follows from the analysis of the 1+1-dimensional case. Then equation
(\ref{bia}) may be equivalently replaced by another equation obtained by applying
$\partial_3$ on it
and using (\ref{arr1})-(\ref{arr11}). This is just the Bianchi identity. We write it
in the form
\begin{equation}\label{arr3} 
{\vec A}_3 \times {\vec B}_3=-\partial_3 {\vec B}_3
-\partial_2 {\vec B}_2 - {\vec A}_2 \times {\vec B}_2 - 
\partial_1 {\vec B}_1 - {\vec A}_1 \times {\vec B}_1
\end{equation}   

Now let us decompose  ${\vec A}_3$ in directions parallel and perpendicular to  
${\vec B}_3$,
\begin{equation}\label{f1}
{\vec A}_3=\alpha {\vec B}_3 + {\vec A}_{3\perp}.
\end{equation}
In the generic case, where $|{\vec B}|\neq 0 $,
equation (\ref{arr3}) determines ${\vec A}_{3\perp}$ entirely.
Taking the cross product of (\ref{arr3}) with ${\vec B}_3$, we get, 
\begin{equation}\label{f4}
{\vec A}_3 = \alpha {\vec B}_3 -\frac{1}{|{\vec B}_3|^2}
{\vec B}_3 \times \left [ ({\vec A}_2 \times {\vec B}_2)
+({\vec A}_1 \times {\vec B}_1) + (\partial_i {\vec B}_i)
\right ].
\end{equation}
where $\alpha$ can be arbitrarily chosen.
 
We now address the question whether $\alpha$ can also be determined.
Taking the dot product of (\ref{arr3}) with ${\vec B}_3 $, we get,
\begin{equation}\label{f12} 
{\vec B}_3 \cdot\partial_i{\vec B}_i+
({\vec B}_3 \times {\vec B}_1)\cdot{\vec A}_1
+({\vec B}_3 \times {\vec B}_2)\cdot{\vec A}_2=0.
\end{equation}
This is a constraint which ${\vec A}_1$ and ${\vec A}_2$ have to satisfy.
It is satisfied on $x_3=0$. In order that it is satisfied at all $x_3$,
we apply $\partial_3$ on (\ref{f12}) and use (\ref{arr2}) and (\ref{ar0}). 
We obtain
\begin{eqnarray}\label{f14}
&-(\partial_1{\vec A}_3-{\vec A}_3\times {\vec A}_1+{\vec B}_2)
\cdot({\vec B}_1\times {\vec B}_3)
-{\vec A}_1\cdot\partial_3({\vec B}_1\times {\vec B}_3) &
\nonumber \\
&-(\partial_2{\vec A}_3+{\vec A}_2\times {\vec A}_3-{\vec B}_1)
\cdot({\vec B}_2\times {\vec B}_3) 
-{\vec A}_2\cdot\partial_3({\vec B}_2\times {\vec B}_3) &
\nonumber \\
&+\partial_3(\partial_i {\vec B}_i)\cdot {\vec B}_3
+(\partial_i {\vec B}_i)\cdot(\partial_3 {\vec B}_3)=0 &
\end{eqnarray}
Now we can substitute the expression for ${\vec A}_3$ from (\ref{f4}).
Note that in this substitution, the derivatives do not act on $\alpha$
since in that case we get terms ${\vec B}_3\cdot{\vec B}_1\times{\vec B}_3$
and ${\vec B}_3\cdot{\vec B}_2\times{\vec B}_3$ which vanish.
We get the coefficient of $\alpha$ as $(D_1[A]{\vec B}_3)\cdot ({\vec B}_1
\times {\vec B}_3) +(D_2[A]{\vec B}_3)\cdot ({\vec B}_2\times {\vec B}_3)$.
Whenever this coefficient is non-zero, the 
linear equation for $\alpha$  is invertible and  this explicitly gives us 
$\alpha$ as a function of ${\vec A}_1, {\vec A}_2$ and $ {\vec B}_i$ .
Generically we do not expect any problem in solving for $\alpha$.  

We now have ${\vec A}_3$ as a local function of ${\vec A}_1, {\vec A}_2$ and $ {\vec 
B}_i$'s and we can
substitute for it in (\ref{arr2}-10). We further expect that 
the field configurations are mostly non-vanishing so that the coefficients $G_{ijr}$ 
and $G_i$ are regular and we can apply the
theorem to get ${\vec A}_1, {\vec A}_2$ and hence ${\vec A}_3$ as unique functionals of
$ {\vec B}_i(x)$. 

Alternatively we could consider the system of equations
\begin{eqnarray}\label{arr5}
\partial_3 {\vec A}_2 &=& \partial_2 {\vec A}_3 + {\vec A}_2 \times {\vec A}_3-
{\vec B}_1 \\
\partial_3 {\vec A}_1 &=& \partial_1 {\vec A}_3 - {\vec A}_3 \times {\vec A}_1+
{\vec B}_2 \label{e2}\\
\partial_3 ({\vec A}_3\times {\vec B}_3) &=& -
(\partial_1{\vec A}_3-{\vec A}_3\times {\vec A}_1+{\vec B}_2)\times {\vec B}_1
\nonumber \\
&&-(\partial_2{\vec A}_3+{\vec A}_2\times {\vec A}_3-{\vec B}_1)\times {\vec B}_2
  \\
&&-{\vec A}_1\times\partial_3{\vec B}_1 - \partial_3(\partial_i {\vec B}_i)
-{\vec A}_2\times\partial_3{\vec B}_2 \nonumber\\
\partial_3 ({\vec A}_3\cdot{\vec B}_3) &=&
\partial_3 (|{\vec B}_3|^2 \alpha({\vec A}_1,{\vec A}_2 , {\vec B}_i)).
\end{eqnarray}
Here in the last equation $\alpha({\vec A}_1,{\vec A}_2 , {\vec B}_i)$ is to be 
replaced by the expression obtained for $\alpha$ from equation (\ref{f14}) and
$\partial_3 {\vec A}_1$ and $\partial_3 {\vec A}_2$ are to be replaced using
(\ref{arr5}) and (\ref{e2}).
This system of equations is in the form where the Cauchy-Kowalevsky theorem can be 
applied and this system uniquely determines all the unknown variables
once the initial data is specified. The first two 
equations contain the six unknowns ${\vec A}_1$ and ${\vec A}_2$. The third one 
contains the two components of ${\vec A}_3$ transverse to ${\vec B}_3$
and the fourth one has the component of ${\vec A}_3$ parallel to ${\vec B}_3$.
Thus all the nine degrees of freedom are uniquely determined.
Therefore generically there are no gauge field copies. The only ambiguity in the
choice of the potential is limited to a subspace which specifies the initial 
conditions as required in the theorem.

\section{Existence of torsion free driebeins for arbitrary A}

In this section we address the question whether there exists any continuous family of 
potentials which generate the same magnetic field. 
Let ${\vec A}_i$ and ${\vec A}_i+\epsilon {\vec e}_i$ 
generate the same magnetic field, where $\epsilon$ is a small parameter.
Then ${\vec e}_i$ satisfies the equation
\begin{equation}\label{f5}
\epsilon_{ijk}(\partial_j {\vec e}_k + {\vec A}_j\times {\vec e}_k )=0.
\end{equation}
This is precisely the equation for a driebein ${\vec e}_i$ to have zero torsion 
with respect to the connection one form ${\vec A}_i$.
Thus we are asking if there exists a driebein with zero torsion for a given arbitrary 
connection one form. This is an important question in the context of general 
relativity.
We also have a consistency condition by taking the covariant derivative of this 
equation. That is given by
\begin{equation}\label{f6}
{\vec B}_k\times {\vec e}_k = 0
\end{equation}
Let us rewrite the equations in a more convenient way. We take our system of 
equations as
\begin{eqnarray}\label{f7}
\partial_3 {\vec e}_2 &=& \partial_2 {\vec e}_3 +{\vec A}_2\times {\vec e}_3
-{\vec A}_3\times {\vec e}_2 \\
\partial_3 {\vec e}_1 &=& \partial_1 {\vec e}_3 +{\vec A}_1\times {\vec e}_3
-{\vec A}_3\times {\vec e}_1 \label{e1}
\end{eqnarray}
and the consistency condition (\ref{f6}). This set is equivalent to the 
set of equations (\ref{f5}).
As in the previous case, we first 
look at the consistency condition. Let us decompose ${\vec e}_3$ as
\begin{equation}\label{f8}
{\vec e}_3= \beta {\vec B}_3 + {\vec e}_{3\perp}
\end{equation}
Again (\ref{f6}) fixes for us ${\vec e}_{3\perp}$ in terms of the magnetic fields
(in the generic case ${\vec B}_3\neq 0$).
We get,
\begin{equation}\label{f21}
{\vec e}_{3}= \beta {\vec B}_3 - \frac{1}{|{\vec B}_3|} {\vec B}_I\times{\vec e}_I
\end{equation}
where $I$ goes over $1,2$.
Now we can substitute this form of ${\vec e}_3$ in the equations
(\ref{f7}-\ref{e1}).
We obtain ${\vec e}_1$ and ${\vec e}_2$ as unique functions of $\beta$ and 
the magnetic fields. However this ${\vec e}_1$ and ${\vec e}_2$ have to satisfy the 
consistency conditions
\begin{equation}\label{f9}
{\vec B}_3\cdot {\vec B}_I \times {\vec e}_I=0
\end{equation}
where again $I$ goes over $1,2$.
Taking $\partial_3$ of equation (\ref{f9}), we get, using
(\ref{f7}) and (\ref{e1})
\begin{equation}\label{f10}
D_3 ({\vec B}_3\times {\vec B}_I) \cdot {\vec e}_I
+{\vec B}_I \cdot {\vec B}_3 \times D_I{\vec e}_3 = 0
\end{equation}
Putting in the expression of ${\vec e}_{3}$, we get a linear equation for $\beta$
\begin{equation}\label{f11}
D_3 ({\vec B}_3\times {\vec B}_I) \cdot {\vec e}_I
+({\vec B}_I \times {\vec B}_3)\cdot (D_I {\vec B}_3)\beta
-({\vec B}_I \times {\vec B}_3)\cdot D_I [\frac{1}{|{\vec B}_3|}
({\vec B}_J \times {\vec e}_J)] = 0
\end{equation}
This equation can be inverted to solve for $\beta$
as a function of ${\vec e}_1,{\vec e}_2,{\vec A}_1,{\vec A}_2$
and ${\vec B}_i$ whenever $({\vec B}_I \times 
{\vec B}_3)\cdot (D_I {\vec B}_3)$ is non-zero.

Formally we could have also looked at the set of equations
\begin{eqnarray}
\partial_3 {\vec e}_2 &=& \partial_2 {\vec e}_3 +{\vec A}_2\times {\vec e}_3
-{\vec A}_3\times {\vec e}_2 \label{ee1}\\
\partial_3 {\vec e}_1 &=& \partial_1 {\vec e}_3 +{\vec A}_1\times {\vec e}_3
-{\vec A}_3\times {\vec e}_1 \label{ee2}\\
\partial_3 ({\vec B}_3\times {\vec e}_3)& =& -(\partial_3{\vec B}_2)\times 
{\vec e}_2 - (\partial_3{\vec B}_1)\times{\vec e}_1 \nonumber \\
&&-{\vec B}_2\times (\partial_2 {\vec e}_3 +{\vec A}_2\times {\vec e}_3
-{\vec A}_3\times {\vec e}_2) \nonumber \\ 
&&- {\vec B}_1\times (\partial_1 {\vec e}_3 +{\vec 
A}_1\times {\vec e}_3 -{\vec A}_3\times {\vec e}_1) \\
\partial_3 ({\vec B}_3\cdot{\vec e}_3)&=&
\partial_3 (|{\vec B}_3|^2)\beta({\vec e}_1,{\vec e}_2,{\vec A}_1,{\vec A}_2,{\vec B}_i) .
\end{eqnarray}
In the last equation, $\beta$ has to be replaced by its solution from (\ref{f11})
and $\partial_3 e_I$ is to be substituted from (\ref{ee1}) and (\ref{ee2}).

We expect the non-Abelian potentials and magnetic fields are smooth and non-vanishing so 
that the coeffiecient functions for the set of differential equations are regular. 
 Applying the Cauchy-Kowalevsky theorem to this set, we 
get a unique smooth solution for ${\vec e}_1, {\vec e}_2$ and ${\vec e}_3$.
 Thus for any potential there is a torsion free driebein, and the only ambiguity is in 
the choice of the driebein to fix the initial conditions required by the theorem. 

\section{An explicit calculation}

We now illustrate these results by an explicit calculation 
 for the special case $A_i^a=\delta_i^a$. 
In  momentum space, the equation looks like 
\begin{equation}
\epsilon_{ijk} (-i p_j \delta^{ac} +\epsilon_{abc} \delta_j^b ) e_k^c (p)=0
\end{equation}
or
\begin{equation}
( -i \epsilon_{ijk} p_j \delta^{ac} + \delta_i^a \delta_k^c - \delta_i^c \delta_k^a )
 e_k^c (p)=0
\end{equation}
In three dimensions we can choose three orthogonal vectors. We choose three such vectors 
as $ ( {\vec p},{\vec n}, {\vec m} ) $ where ${\vec p}$ coincides with the ${\vec p}$ 
which appears in the equation and ${\vec n}$ and ${\vec m}$ are unit vectors . We also 
orient $ ( {\vec p},{\vec n}, {\vec m} ) $ such 
that ${\vec p} \times {\vec m}= |{\vec p}|{\vec n}$ and $ {\vec p} \times {\vec n}= 
- |{\vec p}|{\vec m}$. Next we write a general solution for $ e_k^c$ in terms of the 
dyad basis as
\begin{eqnarray}
&e_{kc} = a_1\, n_c m_k + a_2\, n_k m_c + a_3\, n_k n_c + a_4\, m_k m_c& \nonumber \\ 
&+ a_5\, p_c m_k + a_6\, p_k m_c + a_7\, p_c n_k + a_8\, p_k n_c + a_9\, p_k p_c, &
\end{eqnarray}
where $a_i$'s are unknown coefficients to be determined. 

Substituting the solution in the equation, we get various relations among the coefficients.
$a_5, a_6, a_7, a_8 $ and $a_9$ turn out to be zero identically. In addition we get
\begin{equation}
-i |{\vec p}| a_1 = -i |{\vec p}|^3 a_2 = a_3 = |{\vec p}|^2 a_4.
\end{equation}
Therefore, we get a non-zero solution only if 
\begin{equation}
|{\vec p}|=1,
\end{equation}
 in which case,
\begin{equation}
-i a_1 = -i a_2 = a_3 = a_4=a
\end{equation}
Thus the general solution is
\begin{equation}
e_{ib}(x)=\int\,d\Omega\, a(\Omega) e^{i{\hat p}\cdot x} ({\hat m}+
i{\hat n})_i ({\hat m}-i{\hat n})_b
\end{equation}
Here the integration is over all directions of the vector ${\hat p}$.
The solutions have an arbitrary function $a(\Omega)$. We may fix $a(\Omega)$ by
using initial data on $x_3=0$ surface. This may be interpreted as the arbitrary 
choice of ${\vec e}_i(x)$ at the boundary. However if we require ${\vec e}_i(x)$
vanishes rapidly at infinity, there may not be any solutions. Thus gauge copies would 
be absent in this case. 

A similar exercise can be carried out for any constant vector potential and gives an 
identical result. 

\section{Conclusions}

In this chapter we have looked at two problems regarding the existence of non-Abelian 
vector potentials. First we asked the question if there exists a vector potential 
for any arbitrary magnetic field. 
We found that there are many choices of ${\vec A}_i(x)$ on the $x_3=0$ surface which 
reproduces ${\vec B}_i(x)$ on the surface. (This is the gauge field ambiguity in
1+1 dimensions.) For each such boundary condition on ${\vec A}_i(x)$ we have seen 
(in the generic case) that there is a unique potential ${\vec A}_i(x)$ which reproduces
the given magnetic field everywhere. The non-Abelian Bianchi identity does not 
constrain the non-Abelian magnetic fields in contrast to the abelian case. The 
ambiguity in the choice of the potentials is (in the generic case) only due to the 
ambiguity in ${\vec A}_i(x)$ on the $x_3=0$ surface. Thus it is related to the gauge copy
problem in 1+1 dimensions.

We thank Professors Ramesh Anishetty and K. Mariwalla for helpful 
discussions and Prof. P.P.Divakaran for a useful comment.

\end{document}